\documentclass[10pt]{ismd08}
\usepackage{graphicx}
\usepackage{cite,./mcite}

\begin{document}

\def\eq#1{{Eq.~(\ref{#1})}}
\def\fig#1{{Fig.~\ref{#1}}}

\title{Theory Summary: International Symposium on Multiparticle Dynamics 2008} 

\author{Yuri V.\ Kovchegov}

\institute{Department of Physics, The Ohio State University, Columbus, OH
  43210, USA} \maketitle

\begin{abstract}
  I summarize the theory talks presented at the International
  Symposium on Multiparticle Dynamics 2008.
\end{abstract}

\section{Introduction}
\label{general}

\begin{figure}[t]
  \begin{center}
    \includegraphics[width=14.5cm]{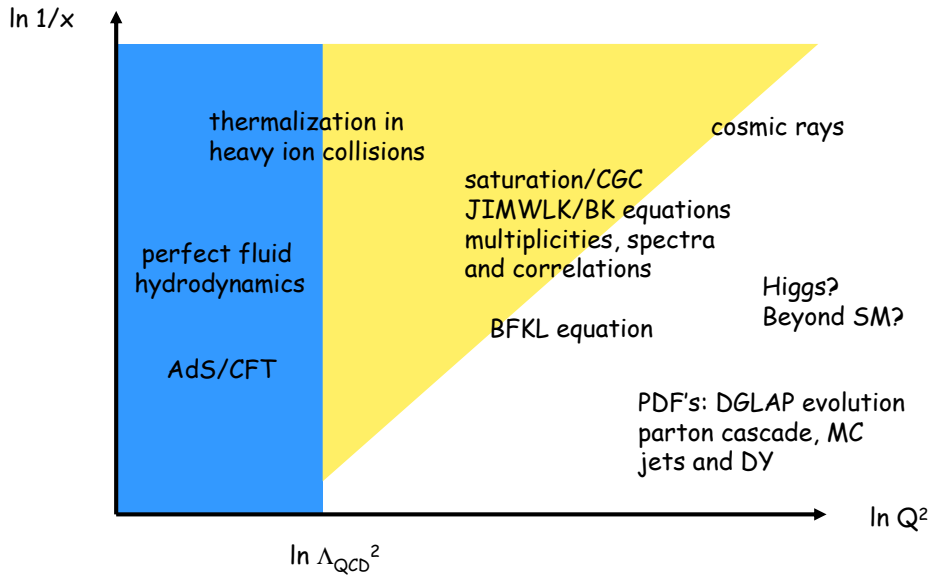}
  \end{center}
\vspace*{-1.5cm}
  \caption{My own version of arranging all the topics covered during the 
    conference in the $(\ln Q^2, \ln 1/x_{Bj})$-plane. The idea was
    borrowed from the conference poster with the topics somewhat
    modified and moved around.  }
  \label{map}
\end{figure}

The XXXVIII International Symposium on Multiparticle Dynamics (ISMD
2008) covered a wide variety of topics in nuclear and particle
physics. The organizers had an interesting idea of arranging all the
topics on a single plot in the $(\ln Q^2, \ln 1/x_{Bj})$-plane, as
shown in the conference poster. I think the idea of such
classification on such a broad scale is new and interesting: in
\fig{map} I present my own version of the classification proposed by
the organizers with some slight modifications as compared to the
original. \fig{map} gives the summary of the topics covered during the
conference: below I will discuss each of the topics shown in \fig{map}
in a separate Section.

Indeed no classification can adequately reflect all the subtleties of
each of the topics discussed.  The classification of \fig{map} is no
exception. Many of the subjects shown have a lot more dimensions to
them (in some cases literally so) than shown in \fig{map}.

The idea of mapping out the $(\ln Q^2, \ln 1/x_{Bj})$-plane comes from
the physics of parton saturation at small Bjorken $x$, also known as
the Color Glass Condensate (CGC) (for a review see
\cite{JalilianMarian:2005jf}). It appears that this approach could be
generalized beyond small-$x$ physics. One of the main concepts of CGC
is that at small enough $x$ the gluon density in the proton or nuclear
wave functions gets so high that non-linear effects, such as parton
mergers, become important leading to saturation of gluon and quark
distribution functions. The transition to this saturation regime is
described by the saturation scale $Q_s$, which is a function of $x$.
$Q_s$ increases as $x$ decreases. Saturation region is schematically
represented by a yellow triangle in \fig{map}. Indeed strong
interaction physics below the confinement scale, at $Q^2 <
\Lambda^2_{\mathrm{QCD}}$, is non-perturbative. The non-perturbative
large-coupling region is denoted in \fig{map} by a blue rectangle. At
small enough $x$ the saturation scale becomes larger than
$\Lambda_{\mathrm{QCD}}$: therefore the saturation regime lies in the
perturbative region to the right of $\Lambda_{\mathrm{QCD}}$.

The large-$Q^2$ region with not very small $x$ is the domain of linear
DGLAP evolution equation
\cite{Dokshitzer:1977sg,Gribov:1972ri,Altarelli:1977zs}. This is the
region where collinear factorization applies. The approaches based on
collinear factorization, such as parton cascade simulations and jet
physics in general, also belong in that region. Some of the topics
discussed in that subfield will be described in Sect. \ref{PDF} below.
As one moves towards smaller $x$ (and somewhat lower $Q^2$) the
logarithms of $1/x$ become important. Such logarithms are resummed by
the BFKL equation \cite{Bal-Lip,Kuraev:1977fs}.  Progress in our
understanding of BFKL will be reviewed in Sect.  \ref{BFKL}. Moving on
toward even lower $x$ one crosses the saturation line and enters the
saturation/CGC region. Here the nonlinear JIMWLK
\cite{Jalilian-Marian:1997gr,Iancu:2000hn} and BK
\cite{Balitsky:1996ub,Kovchegov:1999yj} evolution equations apply. I
have also grouped in this region of the map the predictions of CGC
physics for various $AA$, $pA$ and $pp$ observables.  The talks on
this topic will be discussed in Sect.  \ref{CGC}. All the small-$x$
machinery should be directly applicable to cosmic ray physics: the
progress in this direction will be mentioned in Sect. \ref{Cosmic}.
Heavy ion physics poses a number of important questions for theorists.
Over the past several years a consensus has been reached in the heavy
ion community that heavy ion collisions at RHIC lead to the creation
of a strongly-coupled quark-gluon plasma (QGP). The challenges facing
the heavy ion theory community include understanding of the creation
of such medium: how do the particles produced in a collision
thermalize to form the strongly-coupled QGP? The mechanism leading to
the creation of strongly-coupled QGP may or may not be perturbative,
as reflected in \fig{map}. The talks on this subjects will be reviewed
in Sect.  \ref{therm}. The subsequent evolution of the produced medium
governed by the perfect fluid or viscous hydrodynamics will be
discussed in Sect.  \ref{hydro}. Developments in Anti-de
Sitter/Conformal Field Theory (AdS/CFT) correspondence, which can shed
light on many topics in heavy ion collisions, deep inelastic
scattering (DIS), and hadronic scattering, will be reviewed in Sect.
\ref{ads}. Finally, the Higgs boson and physics beyond Standard Model
are placed at large $Q^2$ and at large energy/small-$x$ in \fig{map}:
they will be mentioned in Sect. \ref{higgs+}.

ISMD 2008 featured a large number of very interesting talks. I have to
apologize beforehand for not being able to cover all of them due to
space limitations. Also, when describing work presented at ISMD 2008 I
will not provide explicit citations to the corresponding publications,
assuming that interested readers could find the needed references in
the Proceedings contributions of the corresponding speakers. Finally,
as this is not a review article, in presenting the topics I will not
spend much time recounting many important successes in each subfield,
but will concentrate instead on open problems at the forefront of
research.

\section{PDF's, parton cascades and jets }
\label{PDF}

\begin{figure}[h]
  \begin{center}
    \includegraphics[width=12cm]{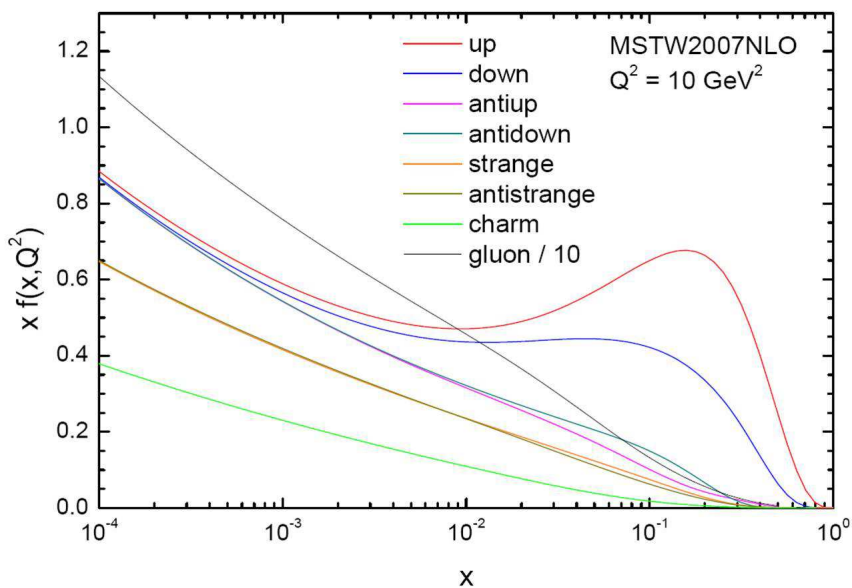}
  \end{center}
\vspace*{-1cm}
  \caption{PDF's in the MSTW 2007 parameterization (from the talk by Stirling).}
  \label{PDFs}
\end{figure}

Much of our knowledge about QCD at high energies comes from and could
be summarized in parton distribution functions (PDF's). Our current
knowledge of PDF's was summarized in the talk by Stirling. \fig{PDFs}
presents the proton PDF's at $Q^2 = 10$~GeV$^2$ given by the MSTW 2007
parameterization.

\begin{figure}
  \begin{center}
    \includegraphics[width=7cm]{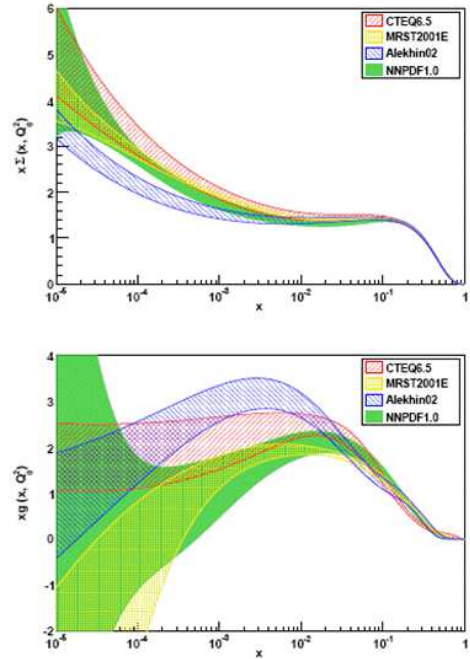}
  \end{center}
\vspace*{-.7cm}
  \caption{Singlet and gluon distribution function at $Q_0^2 = 2$~GeV$^2$ 
    due to different PDF collaborations with the error bars shown
    (from the talk by Rojo-Chacon).}
  \label{err}
\end{figure}

There has been much improvement in our understanding of PDF's in
recent years. Error analysis have been carried out and now many PDF's
come with error bars, as demonstrated in \fig{err} shown in the talk
by Rojo-Chacon. \fig{err} shows singlet and gluon distribution
functions at $Q_0^2 = 2$~GeV$^2$ due to CTEQ, MRST/MSTW, Alekhin and
NNPDF collaborations
\cite{Lai:1999wy,Martin:2007bv,Alekhin:2002fv,Ball:2008by} along with
the error bars. We see that in the small-$x$ region PDF uncertainties
are large. They appear to increase as we go toward lower Bjorken $x$
into the region where there is no data.

The lower panel of \fig{err} also shows that at small-$x$ and
small-$Q^2$ the gluon distribution function becomes {\sl negative}.
This issue had been discussed a lot over the past years, and received
a lot of attention at ISMD 2008 as well. The question is whether a
negative gluon distribution necessarily implies a breakdown of the
approach based on the linear DGLAP evolution equation. The standard
argument against DGLAP breakdown is that at small $Q^2$ the
expectation value of the operator identified with the gluon
distribution function does not anymore count the number of gluons.
Therefore no fundamental law is violated if it goes negative. As was
brought up in the discussion session by Cooper-Sakar, one has to look
at the structure function $F_L$, which is closely related to the gluon
distribution function.  $F_L$ is indeed a physical observable
expressible in terms of scattering cross sections: it has to be
positive. If $F_L$ resulting from the gluon distribution functions in
the lower panel of \fig{err} remains positive, then one could argue
that there is no problem with the negative gluon distribution
function. As indeed the $F_L$'s obtained from the gluon distribution
functions in \fig{err} appear to be positive one indeed can argue that
negative $xG$ are allowed.

To me such arguments sound a bit like epicycles in Ptolemaic
astronomy: some of our colleagues are trying to rescue a theory in
trouble. Strictly-speaking it is true that there is nothing requiring
$xG$ to be positive definite everywhere. However, I spent many years
calculating $xG$ at small-$x$ in the perturbative (saturation)
framework and never saw it go negative. It would be interesting and
convincing if the proponents of negative $xG$ could come up with a
(purely theoretical) model for gluon distribution, where everything is
perturbative and under calculational control, and where $xG$ does
become negative at small-$x$ and small-$Q^2$. For instance one could
study gluon distribution in a very heavy quarkonium. Large quark
masses would insure small coupling allowing to calculate $xG$
perturbatively from first principles. If negativity of $xG$ at low-$x$
and low-$Q^2$ is a natural property of the gluon distribution
operator, it should come out straightforwardly in such a calculation.

Parton cascades as the way to model actual collisions on the
event-by-event basis received a lot of attention at ISMD 2008 as well
with a nice review talk by Z. Nagy. The ideas of going beyond
collinear factorization and including $k_T$-dependent effects into
parton cascades were discussed by Hautmann. Problems with Monte Carlo
simulations of small-$x$ coherent effects were discussed in the talk
by Marchesini. There is a difficult problem that arises when one tries
to include recoil effects into the color-dipole parton cascades in a
probabilistic QCD picture.

\begin{figure}
  \begin{center}
    \includegraphics[width=7cm]{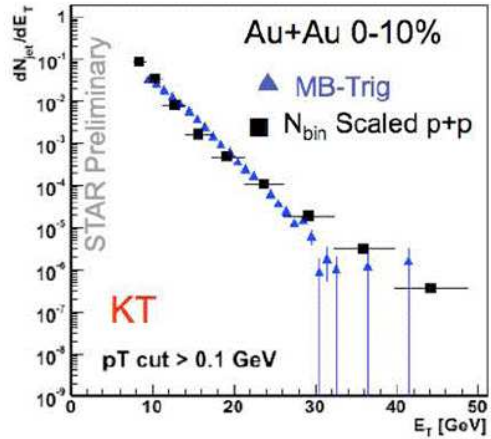}
  \end{center}
\vspace*{-.7cm}
  \caption{Preliminary STAR collaboration data on the number of jets 
    as a function of $E_T$ of the jet for central $Au+Au$ collisions
    (triangles) and for proton-proton collisions scaled up by the
    number of binary collisions (squares). (Picture is from the
    talk by Rojo-Chacon, originally taken from \protect\cite{Salur:2008hs}.)}
  \label{jets}
\end{figure}
There were several good talks on jet analysis and algorithms. I was
particularly interested to see jet analyses coming to RHIC. The
suppression of high-$p_T$ hadrons produced in $Au+Au$ collisions at
RHIC as compared to $p+p$ collisions scaled up by the number of binary
collisions is believed to be one of the smoking guns for the creation
of a hot and dense medium in heavy ion collisions, likely to be a
thermalized quark-gluon plasma (QGP)
\cite{Adler:2003ii,Adams:2003im,Arsene:2003yk}. The suppression is
quantified with the help of the nuclear modification factor $R^{AA}$.
The suppression was observed in RHIC experiments at $\sqrt{s}=200$~GeV
and attributed to parton energy loss also known as jet quenching.
However one should not forget that in the $R^{AA}$ measurements one
measures individual high-$p_T$ hadrons, and not proper jets. A jet
analysis with a jet cone definition has recently been carried out by
the STAR experiment.  The preliminary results are shown in \fig{jets},
which was shown at ISMD 2008 by Rojo-Chacon with a similar figure
shown by Caines.  \fig{jets} depicts the number of jets as a function
of $E_T$ of the jet.  In \fig{jets} the triangles denote the data for
$Au+Au$ collisions, while the squares denote the $p+p$ data scaled up
by the number of binary collisions. It is curious and a bit puzzling
that no visible suppression of jets in $Au+Au$ compared to scaled-up
$p+p$ was found (within error bars). One could speculate that the
energy deposited by the hard parton into the medium is not simply
absorbed by the medium, but instead travels along with the parton in
the form of softer partons, such that the net energy in the jet cone
does not change and the jet as a whole does not get suppressed. Indeed
more work is needed to understand the data in \fig{jets}.

\section{The BFKL equation}
\label{BFKL}

The status of the linear BFKL evolution equation has been reviewed in
the talk by White. The main problem with the linear BFKL evolution is
the large and negative NLO BFKL correction to the pomeron intercept,
which one obtains by using the NLO BFKL results of
\cite{Fadin:1998py,Ciafaloni:1998gs} evaluated at the LO saddle point.
The correction is so large that it makes the gluon distribution
function fall off with decreasing $x$. 
\begin{figure}[h]
  \begin{center}
    \includegraphics[width=12cm]{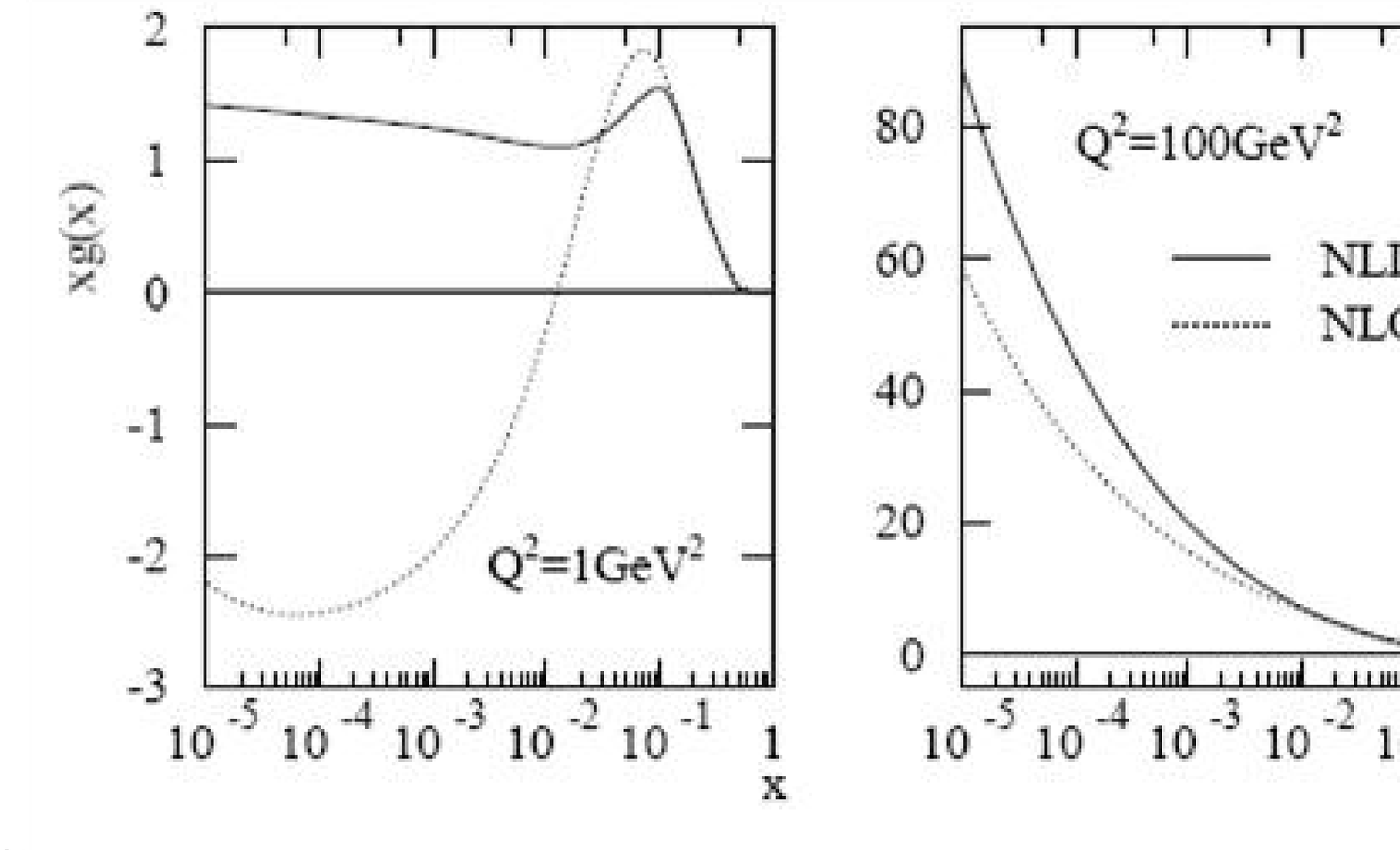}
  \end{center}
\vspace*{-.7cm}
  \caption{Gluon distribution function due to NLO BFKL corrections resummed 
    in the TW prescription (solid lines) compared to the NLO DGLAP
    results (dotted lines). (Pictures are from the talk by
    White.) }
  \label{nlo_bfkl}
\end{figure}

Several ways to remedy this problem have been proposed. It was
observed that going beyond the saddle point approximation, e.g. by
solving the NLO BFKL equation numerically, significantly reduces the
NLO correction to the intercept, making the resulting BFKL Green
function rise with decreasing $x$ \cite{Andersen:2003wy}. An
alternative/complimentary way out involves resumming DGLAP transverse
logarithms in the NLO BFKL kernel: one such procedure, pioneered by
Ciafaloni, Colferai, Salam, and Stasto (CCSS) \cite{Ciafaloni:2003rd}
also gives a positive pomeron intercept albeit somewhat smaller than
the LO BFKL intercept.  Other procedures involved are due to
Altarelli, Ball, and Forte (ABF) \cite{Altarelli:1999vw} and Thorne
and White (TW).  The results of the TW resummation for the gluon
distribution function are shown in \fig{nlo_bfkl} (solid lines)
compared to the NLO DGLAP results (dotted lines). One can see that TW
resummation cures the problem of the negative gluon distribution at
low-$Q^2$ and low-$x$ that NLO DGLAP has. Still the gluon distribution
in the left panel of \fig{nlo_bfkl} corresponding to $Q^2 = 1$~GeV$^2$
is almost flat as one goes toward lower $x$: it is unclear what
physical mechanism would make $xG$ behave in such a way in the absence
of saturation effects in the approach used.

Other problems of the linear BFKL evolution include violation of
unitarity bound (or, more precisely, the black disk limit) and
diffusion into the infrared. Those problems are remedied by the
physics of parton saturation, to be discussed next.

\section{Saturation/Color Glass Condensate}
\label{CGC}

\begin{figure}[h]
  \begin{center}
    \includegraphics[width=10cm]{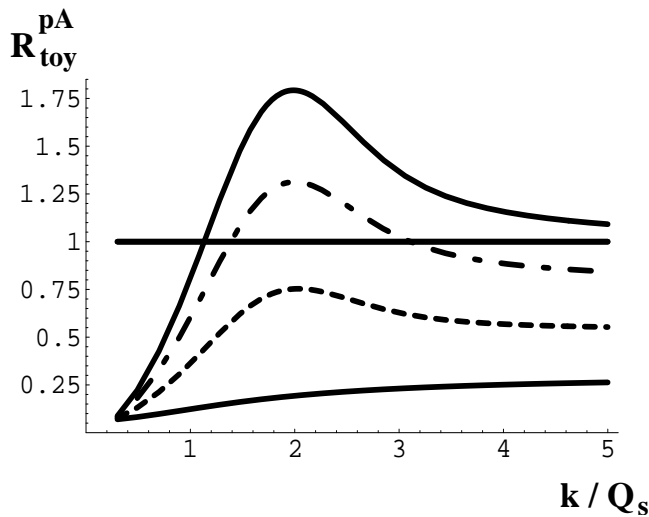}
  \end{center}
\vspace*{-.7cm}
  \caption{A sketch of the nuclear modification factor $R^{pA}$ as a 
    function of the transverse momentum of the produced particle $k_T$
    in the units of the saturation scale from
    \protect\cite{Kharzeev:2003wz}. The upper curve corresponds to the
    lowest energy/rapidity, while the other curves show what happens as
    the energy/rapidity increases.}
  \label{rda}
\end{figure}

The talks by Golec-Biernat and by Marquet gave a nice introduction to
the physics of parton saturation/CGC and the non-linear evolution
equations involved
\cite{Jalilian-Marian:1997gr,Iancu:2000hn,Balitsky:1996ub,Kovchegov:1999yj}.
While the theoretical framework behind CGC is solid, the question of
unique experimental detection of CGC is still debated. CGC prediction
of hadron suppression at forward rapidities in the $d+Au$ collisions
at RHIC \cite{Kharzeev:2002pc,Kharzeev:2003wz,Albacete:2003iq} shown
here in \fig{rda} were spectacularly confirmed by the data
\cite{Arsene:2004ux,Adler:2004eh,Back:2004bq,Adams:2006uz}. The CGC
prediction involved the conventional {\sl shadowing} effects, which
redistribute the partons through multiple rescatterings from lower
$p_T$ to higher $p_T$, leading to low-$p_T$ suppression (shadowing)
and high-$p_T$ enhancement (anti-shadowing) shown in the upper curve
in \fig{rda}. (The high-$p_T$ enhancement of produced particles is
known as Cronin effect.) The effects of small-$x$ BFKL/JIMWLK/BK
evolution equations (the {\sl saturation} effects) lead to decrease of
the number of produced particles (as compared to the $p+p$ reference)
at all $p_T$, as shown by the dash-dotted, dashed, and the lower solid
curves in \fig{rda} (for a review see \cite{JalilianMarian:2005jf}).

\begin{figure}[ht]
  \begin{center}
    \includegraphics[width=10cm]{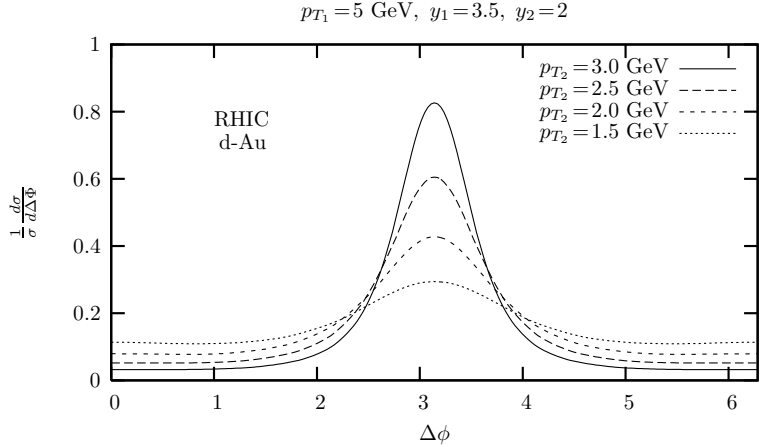}
  \end{center}
\vspace*{-.7cm}
  \caption{CGC prediction for the azimuthal dependence of the two-hadron 
    correlation function for different values of the $p_T$ of the
    associate particle (from the talk by Marquet).}
  \label{corrM}
\end{figure}

However, as conventional approaches based on collinear factorization
with significantly {\sl ad hoc} modified nuclear shadowing have been
able to describe the data {\sl a posteriori} \cite{Eskola:2008ca}, the
need arose for new experimental tests to uniquely disentangle between
the collinear factorization scenario with shadowing included and the
physics of CGC.  One of such CGC predictions for a two-particle
correlation function was shown by Marquet and is presented here in
\fig{corrM}, which shows a two-hadron correlation function plotted
versus the opening azimuthal angle between the two hadrons $\Delta
\phi$. The trigger particle has rapidity $y_1 = 3.5$ and $p_{T 1} =
5$~GeV.
The associate particle has rapidity $y_2 = 2$. The transverse momentum
of the associate particle $p_{T 2}$ is different for different curves,
as explained in \fig{corrM}. The CGC prediction is that as $p_{T 2}$
gets lower and comes closer to the saturation scale $Q_s$ (which is of
the order of $1-2$~GeV at RHIC), the saturation effects would ``wash
out'' the back-to-back azimuthal correlations, leading to a decrease
in the correlation function as predicted in \fig{corrM}. The
experiments currently running at RHIC will be able to test this
prediction.

\begin{figure}[ht]
  \begin{center}
    \includegraphics[width=9cm]{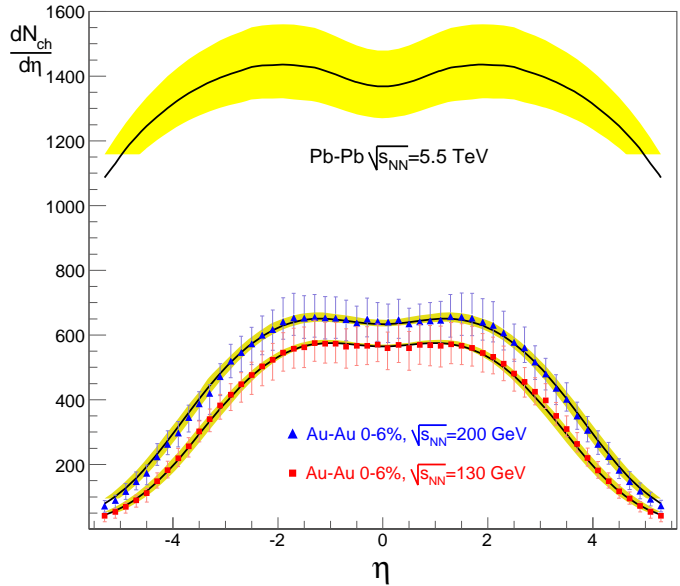}
  \end{center}
\vspace*{-.7cm}
  \caption{CGC prediction from \protect\cite{Albacete:2007sm} for the 
    total charged hadron multiplicity in $Pb+Pb$ collisions at LHC,
    along with successful fits of the same quantity measured by RHIC
    at two different center-of-mass energies. The yellow band around
    the LHC prediction indicates the error bars.}
  \label{dndy}
\end{figure}

Another test of CGC will come from the upcoming LHC heavy ion
experiments. In heavy ion collisions it is hard to construct a
rigorous CGC prediction, as the problem of particle production in CGC
for the collision of two nuclei have not been solved analytically. One
therefore constructs models based on $k_T$-factorization formula
(proven for $p+A$ collisions in CGC in \cite{Kovchegov:2001sc} but not
proven for $A+A$) trying to mimic as close as possible the true CGC
physics (see e.g. \cite{Kharzeev:2001yq}). One of the less
model-dependent predictions of such an approach is for the total charged
hadron multiplicity in heavy ion collisions. Predictions for total
charged hadron multiplicity in $Pb+Pb$ collisions at the LHC from the
work of Albacete \cite{Albacete:2007sm} were shown by Marquet and are
reproduced here in \fig{dndy}.
The plot in \fig{dndy} resulted from using the $k_T$-factorization
formula also used in \cite{Kharzeev:2001yq}. However, the dipole
scattering amplitudes which enter that formula were evolved using the
BK evolution equation with running coupling corrections, which have
been recently calculated in
\cite{Kovchegov:2006vj,Balitsky:2006wa,Albacete:2007yr}. Thus at least
one of the ingredients used in arriving at \fig{dndy} comes form a
fairly rigorous CGC analysis, which has became available very recently
and never has been used before. Based on that I believe that the
prediction in \fig{dndy} is the best theoretically-founded one.
Unfortunately, due to limitations of our understanding of CGC
mentioned above (concerning the applicability of the
$k_T$-factorization formula to nucleus-nucleus collisions), the
prediction in \fig{dndy} still involves some degree of modeling that
we can not control, and should thus be still taken with care.

\begin{figure}[hh]
  \begin{center}
    \includegraphics[width=14.5cm]{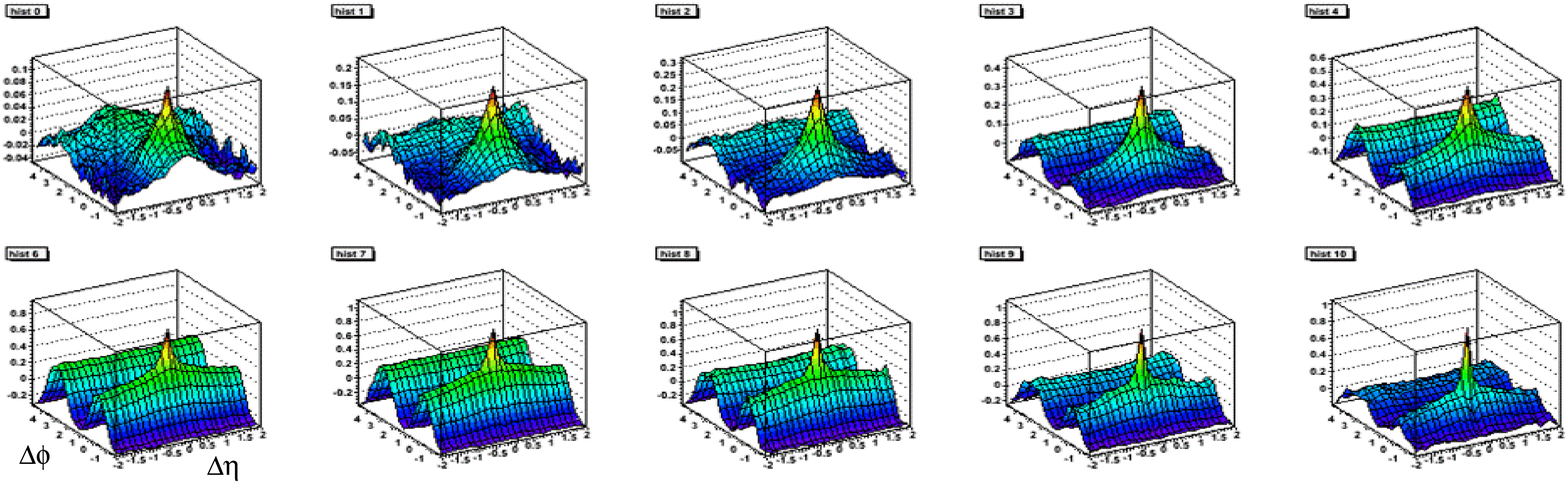}
  \end{center}
\vspace*{-.7cm}
  \caption{Two-hadron correlation function measured in $Au+Au$ collisions 
    at RHIC as a function of rapidity interval between the two hadrons
    $\Delta \eta$ and the azimuthal angle between them $\Delta \phi$.
    Each panel corresponds to a different centrality of the collision.
    The data are from the STAR collaboration
    \protect\cite{Daugherity:2008su}.}
  \label{Corr}
\end{figure}
RHIC experiments continue to surprise us with an amazing quantity of
interesting results. We now know the two-hadron correlation function
as a function of both azimuthal angle between the hadrons and the
rapidity interval between them, as shown in \fig{Corr}. The
correlation function in \fig{Corr} has at least one interesting
unexplained feature: it has long-range rapidity correlations on the
same azimuthal side ($\Delta \phi \approx 0$), known as ``the ridge''.
While many explanations were proposed, the feature remains largely
unexplained. At ISMD 2008 McLerran proposed that ``the ridge'' could
be due to long-range rapidity correlations inherent to CGC. Indeed CGC
predicts rapidity correlations over the intervals of the order of
$\Delta y \sim 1/\alpha_s$, which could be large if the strong
coupling constant $\alpha_s$ is small. The radial flow would then
boost the correlations, confining them to small azimuthal opening
angles, thus creating a ridge-like structure. This is indeed a
plausible explanation, but I feel more detailed theoretical work is
needed to determine whether this is a unique prediction of CGC.

\begin{figure}[h]
  \begin{center}
    \includegraphics[width=8cm]{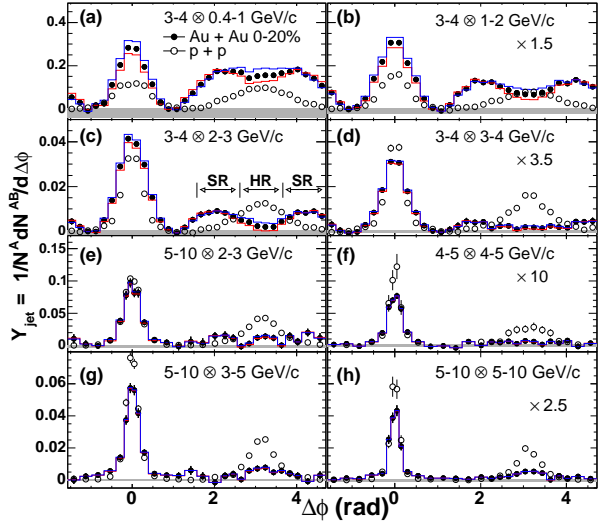}
  \end{center}
\vspace*{-.7cm}
  \caption{Two-hadron azimuthal correlation function as measured by 
    PHENIX experiment (taken from \protect\cite{Adare:2007vu}).}
  \label{mach}
\end{figure}

Another important feature of the two-hadron correlation function in
$Au+Au$ collisions at RHIC is the double-hump structure shown in
\fig{mach}. \fig{mach} depicts the two-particle correlation function
measured by PHENIX collaboration plotted as a function of the
azimuthal angle between the two hadrons.  As one can see from
\fig{mach} the distribution of the associate particles as a function
of azimuthal angle at low transverse momentum of the associate
particle has two maxima. Assuming that the trigger particle travels
through a relatively thin medium layer, one concludes that the
associate particle is likely to travel through a thicker layer of the
medium. The double-hump structure could therefore be caused by a Mach
cone produced by a particle moving through a strongly-coupled medium
\cite{CasalderreySolana:2004qm}. Alternative explanation could be due
to non-Abelian (QCD) Cherenkov radiation, as discussed in the talk by
Dremin. To describe such radiation one has to solve classical
Yang-Mills equations in a medium with some dielectric tensor. (While
indeed Cherenkov radiation is a medium effect, the methods applied to
the analysis are those of CGC, and hence I placed it in the CGC
section.) Cherenkov radiation allows one to describe both STAR and
PHENIX azimuthal correlations data by an appropriate choice of the
dielectric tensor in the medium.

A possible signal of the creation of QGP in heavy ion collisions is
the mass shift for the produced mesons due to medium effects. Padula
suggested that a cleaner way to measure the shift would be by studying
two-particle correlations of $\phi\phi$ and $K^+ K^-$ pairs. Presence
of the mass shift will be signaled by the appearance of back-to-back
correlations in the $\phi\phi$ and $K^+ K^-$ correlators.

Another interesting CGC prediction is for the rapidity distribution of
the net baryon number produced in heavy ion collision. In the talk by
Wolschin it was shown how CGC ideas allow one to successfully describe
baryon number rapidity distribution at SPS and RHIC, and to even make
predictions for LHC. It would be really interesting and important to
measure this quantity at LHC.

A sign of the fact that CGC physics is entering a new era is the
construction of event generators based on CGC concepts and ideas. In
the talks by Avsar and Kutak we have heard about event generators
using CCFM evolution equation with an infrared cutoff mimicking
saturation/CGC effects, similar to how one can mimic the BK equation
by using the BFKL equation with an infrared cutoff. Interesting
results and fits were shown in those talks.

\section{Cosmic rays}
\label{Cosmic}

Ultra-high energy cosmic ray data and the accompanying theory was
presented in the talks by Ostapchenko and Pierog. It was suggested
that the existing discrepancy between the cosmic ray data and the
conventional models may be explained by saturation/CGC effects.  This
is indeed an exciting prospect which needs to be pursued by CGC
practitioners. The progress in this direction can however be marred by
the fact that when extrapolating from current collider energies to the
cosmic ray energies a large uncertainty arises due to the
non-perturbative diffusion of the black disk. As was argued in
\cite{Kovner:2002yt} in perturbative CGC approach the diffusion of
black disk at high energies is very fast: the radius of the disk grows
as a power of energy due to the lack of a mass gap in perturbative
approaches. Any attempt to non-perturbatively model the diffusion by
introducing a mass gap into the problem leads to the radius of the
black disk growing as a logarithm of energy. As the non-perturbative
models are not under the same degree of theoretical control as the
perturbative CGC calculations, the potential theoretical uncertainty
associated with extrapolation to cosmic ray energies could be very
large, leading to uncertainty both in total scattering cross sections
and particle production cross sections calculated in CGC.

\section{Thermalization in heavy ion collisions}
\label{therm}

Understanding the mechanism of thermalization and isotropization of
the medium produced in heavy ion collisions is a very important open
problem in the field. 
\begin{figure}[h]
  \begin{center}
    \includegraphics[width=8cm]{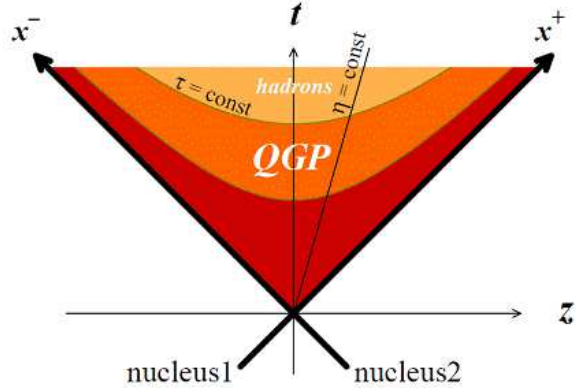}
  \end{center}
\vspace*{-.7cm}
  \caption{Space-time structure of a heavy ion collision (from the talk by Itakura).}
  \label{AA}
\end{figure}
The space-time structure of a heavy ion collision is depicted in
\fig{AA}. The time immediately after the collision is dominated by
particle production. In that region CGC applies, such that the
production of particles is described by perturbative CGC techniques.
This stage of the evolution of the medium is sometimes referred to as
``Glasma'' \cite{Lappi:2006fp}, the term which combines the Color
Glass physics (``Glas'') with the creation of quark-gluon plasma at
later stages of the collision (``lasma''), as shown in \fig{AA}.
However the CGC dynamics by itself leads to a very anisotropic
distribution of the produced matter in momentum space: the end result
of CGC dynamics is a free-streaming ``medium'' with zero longitudinal
pressure component.  Indeed a thermalized medium should have all
pressure components (transverse and longitudinal) equal, as it should
be {\sl isotropic}.  Hence Color Glass itself does not lead to
thermalization, or, more importantly, isotropization of the produced
medium. (Isotropization is a necessary, but not a sufficient condition
of thermalization.)

So how does Color Glass turn into a Glasma? One of the more popular
scenarios was presented in the talk by Itakura and involves magnetic
instabilities in the produced medium (see also
\cite{Mrowczynski:1993qm,Arnold:2003rq}). The main physical idea is
that the momentum space anisotropy of the medium produced in heavy ion
collisions would generate instabilities, which would rapidly
isotropize the system leading to a hydrodynamic behavior of the
medium. This indeed is a plausible scenario of
thermalization/isotropization.

Since the saturation/Color Glass approach gives us a consistent
framework in which all diagrams can be classified and resummed
order-by-order, it is not at all clear why one has to separate the
perturbative dynamics into a part which is incorporated in CGC and
into everything else. From this standpoint the dynamics of
instabilities would correspond to some higher order (quantum)
corrections to the diagrams which we already know how to resum in the
CGC approach. Such corrections would also be a part of CGC, just at
some higher order.  The magnetic instabilities could then be viewed as
some higher order corrections to CGC which somehow got ``out of
control'' and became very large (infinite?). Frankly I am skeptical whether
such corrections exist: all our experience calculating CGC diagrams,
both in the classical framework
\cite{Kovner:1995ja,Kovchegov:1997ke,Krasnitz:2003jw} and including
(LO and NLO) quantum evolution and running coupling corrections
\cite{Kovchegov:2001sc,Kovchegov:2007vf}, never led to any
uncontrolled infinities which would dominate the resulting production
cross sections and the energy-momentum tensor. Perhaps the proponents
of the instability-driven scenario should identify and resum diagrams
with instabilities (starting from the very collision of two nuclei),
and show that their contributions are really important (numerically or
parametrically) and that these diagrams do lead to isotropization of
the medium at late times.  Implications of such diagrams on what we
know in the standard perturbation theory in, say, proton-proton
collisions would also have to be understood. One should also identify
what those new instability diagrams have that was absent in the
multitude of quantum corrections to the classical picture calculated
over the years \cite{Kovchegov:2001sc,Kovchegov:2007vf}.

Alternatively, as the medium created at RHIC is believed to be
strongly-coupled, it is possible that thermalization and
isotropization in heavy ion collisions are essentially
non-perturbative (large-coupling) effects. Such a scenario can not be
quantified in a controlled manner in QCD. However, AdS/CFT
correspondence \cite{Maldacena:1997re,Witten:1998qj} allows us to try
to analyze this problem for the ${\cal N} =4$ Super Yang-Mills (SYM)
theory. In my talk in the Dense Systems session I have presented one
of the efforts in this direction. One could model a heavy ion
collision as a collision of two shock waves in AdS$_5$. Solving
Einstein equations in AdS$_5$ one can find the energy-momentum tensor
of the resulting medium. It has been argued in \cite{Janik:2005zt}
using AdS/CFT correspondence that if one assumes that the produced
medium distribution is rapidity-independent, the strong-coupling
dynamics would inevitably lead at late proper times to an isotropic
medium described by Bjorken hydrodynamics \cite{Bjorken:1982qr}.
However, it is not yet clear whether such a rapidity-independent
distribution would result from a collision of two shock waves. Our
result was that in a strongly-coupled theory the colliding shock waves
would stop shortly after the collision. This seems like a natural
result of the strong coupling effects. If the coupling is strong
enough to stop the colliding nuclei, it is likely to quickly
thermalize the system.  However, a thermal system resulting from
stopping of the nuclei is more likely to be described by
rapidity-dependent Landau hydrodynamics \cite{Landau:1953gs}, instead
of the rapidity-independent Bjorken one. Hence the strong-coupling
effects, if dominant throughout the collision, would not lead to
Bjorken hydrodynamics. On top of that we know from the RHIC data on
net baryon rapidity distribution that valence quarks in the nuclei do
not stop in the collision, and instead (mostly) continue moving along
the beam line \cite{Bearden:2003hx}. Indeed the early stages of the
collisions have to be described by the weak coupling effects, and are
thus outside of the realm of the AdS/CFT correspondence. We presented
a way of mimicking these weak-coupling effects in the dual AdS
geometry.  However, the question of what leads to Bjorken
hydrodynamics still remains open.

\section{Hydrodynamics}
\label{hydro}

Regardless of our lack of understanding of thermalization in heavy ion
collisions, the success of perfect-fluid hydrodynamics description of
particle spectra and elliptic flow $v_2$ measured in the $Au+Au$
collisions at RHIC \cite{Huovinen:2001cy,Teaney:2001av} allows us to
conclude that the medium created in the collisions is probably
strongly coupled and that a hydrodynamic description of such medium is
adequate.
\begin{figure}[h]
  \begin{center}
    \includegraphics[width=10cm]{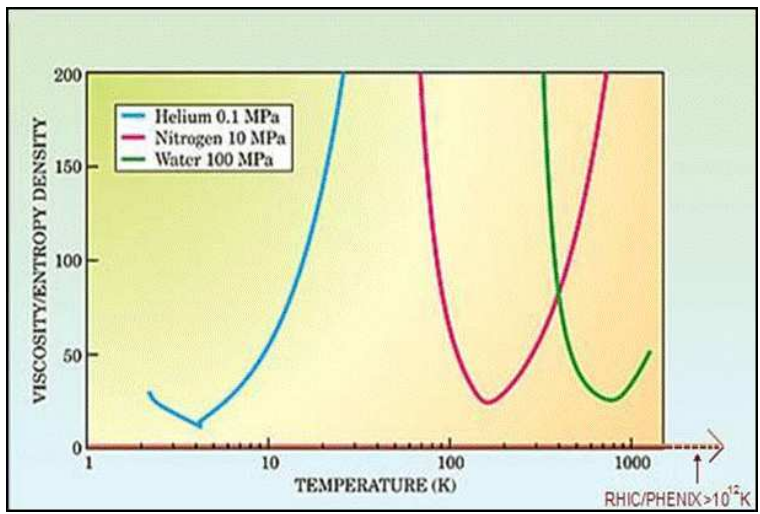}
  \end{center}
\vspace*{-.7cm}
  \caption{The shear viscosity-to-entropy density ratio (in units of $1/4\pi$)
for various media 
    (from the talk by Cs\"{o}rg\H{o}).}
  \label{e/s}
\end{figure}

The Kovtun, Son, Starinets, Policastro (KPSS)
\cite{Policastro:2001yc,Kovtun:2004de} lower bound on the ratio of
shear viscosity $\eta$ to the entropy density $S$ derived from AdS/CFT
correspondence was discussed in the talk by Cs\"{o}rg\H{o}. The KPSS bound
postulates that for any medium (or, more precisely, for any theory
with a gravity dual) one has $\eta/S \ge 1/4 \pi$. \fig{e/s} shows the
ratio of $\eta/S$ plotted as a function of temperature for several
different media with the KPSS bound shown by a straight horizontal
line at the bottom. Cs\"{o}rg\H{o} pointed out that as RHIC data is
consistent with a very low value of $\eta/S$, it is likely that RHIC
fluid is more perfect than any other known fluid. This superfluidity
also takes place at an extremely high temperature, characteristic of
the QGP. However it is still possible that RHIC data allows for higher
values of $\eta/S$ than $1/4 \pi$: by varying the initial conditions
for hydrodynamics one can accommodate somewhat larger values of
$\eta/S$, though the exact values are still under investigation. There
have also been some recent results in string theory suggesting that
the KPSS bound might be violated in some theories due to stringy
(mostly $1/N_c$) corrections. Regardless of that, the low viscosity of the
RHIC QGP still strongly suggests that the medium created in the
collisions is strongly-coupled.

Hydrodynamics is an exciting subfield by itself, allowing for many
interesting exact solutions describing possible evolutions of RHIC
fireball. Many of those solutions have been reviewed in the talk by M.
Nagy, and fall into two main categories: relativistic and
non-relativistic ones.

\begin{figure}[t]
  \begin{center}
    \includegraphics[width=14cm]{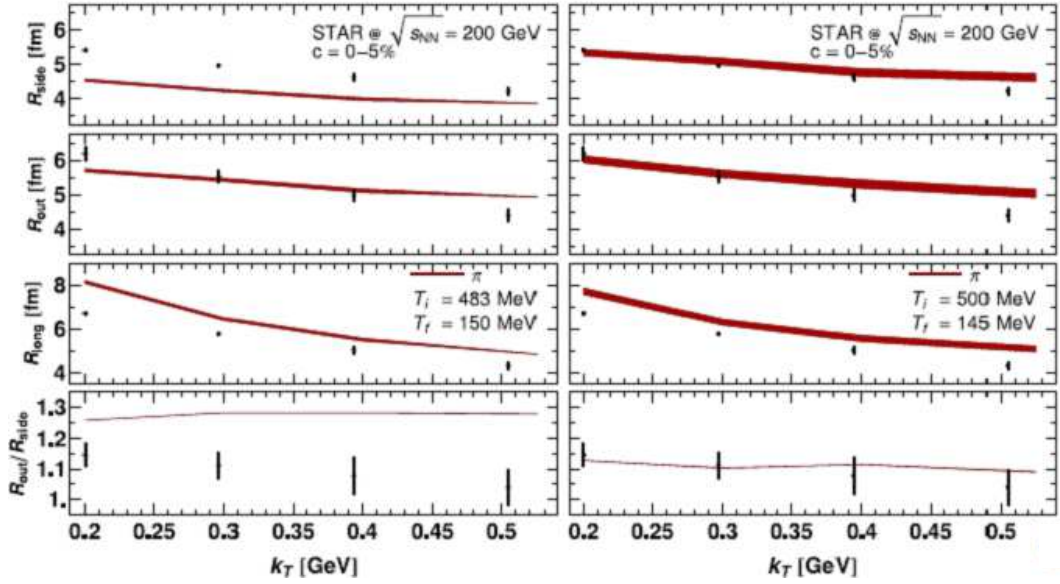}
  \end{center}
  \caption{HBT radii at RHIC compared to hydrodynamic simulations with 
    the standard (Glauber) initial conditions (left panels) and the
    Glauber initial conditions proposed by Florkowski (right panels).
    (The picture is taken from the talk by Florkowski.)}
  \label{HBT}
\end{figure}

There are still open problems with the hydrodynamic description of the
medium produced in RHIC collisions. One is the early thermalization
proper time of $\tau_0 = 0.3 \div 0.5$~fm/c required for hydrodynamics
to describe the data: this problem is related to our (lack of)
understanding of thermalization/isotropization in heavy ion
collisions. Another problem concerns the HBT radii.  While
hydrodynamics is successful in describing particle spectra and $v_2$
\cite{Huovinen:2001cy,Teaney:2001av}, it has been having problems
describing HBT radii. This has been known as the RHIC HBT puzzle (see
e.g.  \cite{Lisa:2008gf}). At ISMD 2008 Florkowski suggested that one
could modify the standard Glauber initial conditions for hydrodynamics
simulations: he suggested starting the simulations with a smaller
Gaussian source, which would generate faster initial expansion.
Apparently this approach worked, allowing to describe the HBT radii,
as shown in \fig{HBT}, using a rather small set of free parameters.
The obtained value for one of the parameters, the thermalization time
$\tau_0$, turns out to be $\tau_0 = 0.25$~fm/c, which is rather close
to some recent estimates based on AdS/CFT approaches
\cite{Kovchegov:2007pq}.

While the approach presented by Florkowski works very well, as can be
clearly seen from \fig{HBT}, one may worry that the initial size of
the Gaussian fireball used is rather small to adequately describe
realistic $Au+Au$ collisions. Therefore in my opinion the conclusion
one can draw from the Gaussian initial conditions analysis is that in
the simulations of \fig{HBT} it mimics some initial time dynamics
which leads to hydrodynamics being initialized with a pretty strong
radial flow. It appears then that in order to describe the HBT radii
one needs the initial conditions for hydrodynamics simulations to
contain large flow in them.  The exact nature of such initial dynamics
still needs to be identified: it might be given by the CGC physics.

The perfect fluid hydrodynamics appears to do a good job at RHIC. It
is possible though that in heavy ion collisions at LHC the plasma that
will be created will start out at higher temperature. This would lead
to smaller coupling constant, thus possibly making the resulting
plasma less strongly coupled. The viscous corrections in such case
would get larger: one therefore needs to construct viscous
hydrodynamics simulations to describe the dynamics of the medium to be
produced in heavy ion collisions at the LHC. But what if viscous
corrections are not enough? What if higher fluid velocity gradients
would also become important? The dynamics of strongly coupled medium
described by AdS/CFT correspondence contains the exact result,
including all gradients of fluid velocity. While obtaining this exact
solution from AdS/CFT correspondence appears to be rather complicated,
one could calculate the viscosity and higher order coefficients in
fluid velocity gradient expansion using AdS/CFT approach.  The results
of the project to calculate the coefficients needed to construct
causal viscous hydrodynamics using the AdS/CFT correspondence were
presented in the talk by Baier. The obtained coefficients could be
used to construct strong-coupling predictions for LHC.

\section{AdS/CFT correspondence}
\label{ads}

AdS/CFT correspondence \cite{Maldacena:1997re,Witten:1998qj} is a very
powerful new tool for studying non-perturbative aspects of gauge
theories coming from string theory (for a review see
\cite{Aharony:1999ti}). AdS/CFT correspondence
\cite{Maldacena:1997re,Witten:1998qj} postulates a duality between the
${\cal N}=4$ SYM theory in 4 space-time dimensions and the type-IIB
string theory in AdS$_5 \times$S$^5$. The more widely used and better
tested gauge-gravity duality suggests that ${\cal N}=4$ SYM theory in
the large-$N_c$ large-$\lambda = g^2 \, N_c$ limit is dual to
classical super-gravity on AdS$_5$ ($\lambda$ is 't Hooft's coupling
constant, $g$ is the gauge coupling). What this means practically is
that in order to find expectation values of various operators in
${\cal N}=4$ SYM theory at large $N_c$ and $\lambda$ one has to
perform classical super-gravity calculations in a curved 5-dimensional
space-time.

A number of talks at ISMD 2008 used the methods of AdS/CFT: some of
these talks I have already mentioned in other Sections. 

A talk by Iancu addressed the question of deep inelastic scattering on
a thermal medium (plasma). In AdS such medium is modeled by the black
brane metric. In the absence of bound states in a conformal theory,
a thermal medium provides a fine target to scatter on. Iancu suggested
calculating a correlator of two R-currents in order to find the
structure functions of the plasma. One of the important results is
that DIS at strong coupling also exhibits the feature of parton
saturation, just like the weakly coupled CGC. The saturation scale in
the theory at strong coupling was found to be equal to $Q_s \sim L \,
T^2$, with $L$ the part of the distance separating the two R-currents
immersed in the medium and $T$ the temperature of the medium. If the
two currents are inside the medium, then $L$ is the distance
separating the currents, and for DIS $L \sim 1/(x T)$ with $x$ the
Bjorken $x$ variable. This gives $Q_s \sim T/x$, i.e. the saturation
scale would grow very strongly as Bjorken $x$ decreases.

\begin{figure}[t]
  \begin{center}
    \includegraphics[width=8cm]{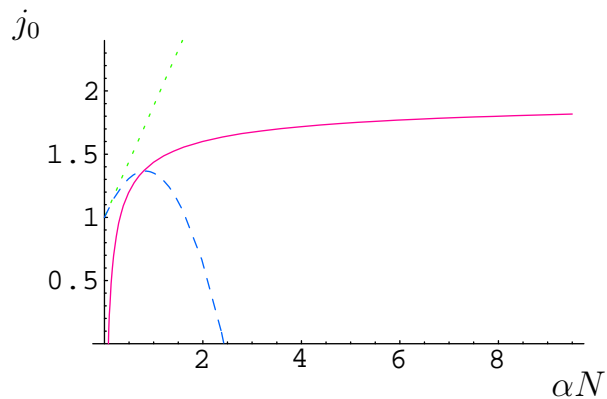}
  \end{center}
\vspace*{-.5cm}
  \caption{ The pomeron intercept $j_0$ in ${\cal N} =4$ SYM as a function 
    of the coupling $\alpha \, N$. The dotted line represents the
    perturbative LO BFKL pomeron intercept, the dashed line is the
    LO+NLO BFKL intercept, and the solid line is the strong-coupling
    AdS result.  (The picture is taken from the talk by Tan.)}
  \label{intercept}
\end{figure}

However, in a realistic high energy DIS scattering the incoming
virtual photon splits into a quark--anti-quark pair very much in
advance of the system hitting the proton or nuclear target. Hence a
more realistic scenario would involve a finite-size medium, such that
$L = 2 R$ with $R$ the radius of the target proton/nucleus. Then one
gets $Q_s \sim R \, T^2$, i.e. the saturation scale is independent of
Bjorken $x$, or, equivalently, of energy.  In this regime the
conclusions presented by Iancu agree with the results of other groups
\cite{Albacete:2008ze,Dominguez:2008vd}. It is rather interesting to
observe that at large coupling the saturation scale becomes
independent of energy. It seems that the classical super-gravity gives
results similar to those given by the classical Yang-Mills fields in
the McLerran-Venugopalan (MV) model \cite{McLerran:1993ni}: there the
saturation scale is also independent of energy. In the MV model we know
that quantum corrections lead to energy-dependence of $Q_s$
\cite{Balitsky:1996ub,Kovchegov:1999yj}. It is possible that quantum
(order $1/\sqrt{\lambda}$) corrections in AdS would make the
saturation scale energy-dependent at strong coupling.

AdS/CFT correspondence allows one to try to understand other related
quantities, such as the intercept of the pomeron and the pomeron
trajectory. The results of such investigations were presented by Tan.
He explained how an AdS/CFT calculation gives the pomeron intercept
$j_0 = 2 - \frac{2}{\sqrt{\lambda}}$ for a strongly-coupled ${\cal N}
=4$ SYM theory. His results are summarized in \fig{intercept}, where
the intercept is plotted as a function of the gauge coupling $\alpha$
times the number of colors in the theory $N$. The dotted and dashed
lines represent the perturbative LO and LO+NLO BFKL intercepts
correspondingly. One can see that the NLO BFKL correction is indeed
large and threatens to make the intercept less than 1 at not very
large $\alpha \, N$. The solid line in \fig{intercept} represents the
AdS strong-coupling result of $j_0 = 2 - \frac{2}{\sqrt{\lambda}}$:
the picture suggests that an interpolation between the two results is
possible, leading to an intercept which is greater than 1 at all
values of the coupling.

At the same time I have to point out that the result of a recent AdS
investigation \cite{Albacete:2008ze} suggests that at high energies
multiple exchanges of the intercept-2 pomerons lead to a somewhat
unphysical behavior of the cross section and violate the black disk
limit. In \cite{Albacete:2008ze} an alternative solution was proposed
with the strong coupling pomeron having an intercept of $j_0 = 1.5$
and with multiple exchanges of such a pomeron giving cross sections
which are unitary and do not violate the black disk limit. More
investigations may be needed to understand the differences between the
two results.

Lipatov talked about another important result related to AdS/CFT
correspondence --- the BDS amplitude ansatz \cite{Bern:2005iz}. He has
argued that the ansatz is violated in the Regge limit, when one
calculates the diagrams contributing to the BFKL evolution. The
violation is relatively minor and only manifests itself in some
channels. This allows one to hope that a modification of the BDS
ansatz is possible which would take into account the discrepancy
presented by Lipatov.

\section{Higgs boson and physics beyond the Standard Model}
\label{higgs+}

LHC had turned on just before the start of ISMD 2008, but had to be
shut down soon after due to a malfunction of the superconducting
magnets. Nevertheless, despite the delay, LHC era is upon us and a
number of talks at ISMD 2008 were dedicated to what one could discover
at LHC. While some aspects of the LHC heavy ion program have been
mentioned above, here I will concentrate on the search for new
particles in proton-proton collisions.

\begin{figure}[h]
\vspace*{-.3cm}
  \begin{center}
    \includegraphics[width=10cm]{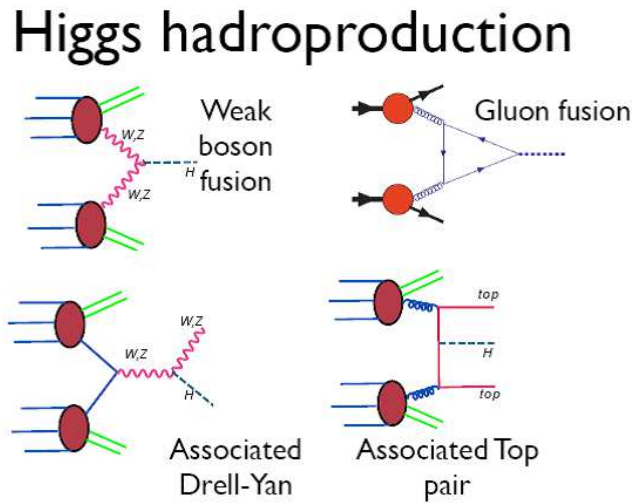}
  \end{center}
\vspace*{-.5cm}
  \caption{Various possible channels of Higgs boson production at LHC 
    (from the talk by Anastasiou.)}
  \label{higgs_pro}
\end{figure}

First of all, if the Standard Model is correct, one expects to be able
to find the Higgs boson at the LHC. Anastasiou gave a talk reviewing
various channels of Higgs production, which are demonstrated in
\fig{higgs_pro}. Hopefully many (or at least one) of these channels
would be observed at LHC.

\begin{figure}[h]
  \begin{center}
    \includegraphics[width=6cm]{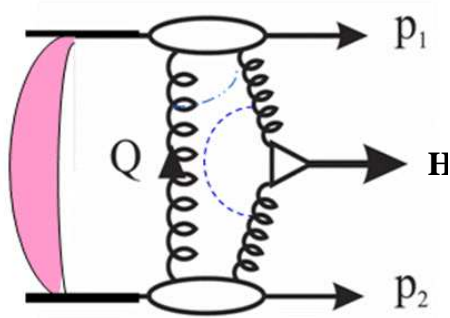}
  \end{center}
\vspace*{-.5cm}
  \caption{Double diffractive Higgs boson production mechanism at the LHC
    (from the talk by V.A. Khoze.)}
  \label{ddhiggs}
\end{figure}

It is possible however that backgrounds at LHC would be too high
making the events shown in \fig{higgs_pro} hard to detect. In this
case a possible cleaner signature of the Higgs would be the double
diffraction production process illustrated in \fig{ddhiggs}, which was
discussed in the talks by Kaidalov and V. A. Khoze. In such process
there will be rapidity gaps between the produced Higgs boson and each
of the protons, allowing for a clean detection of the products of the
Higgs boson decay, and thus for an unambiguous identification of the
Higgs boson.

Unfortunately, as often happens when the soft QCD interactions are
involved, theoretical predictions for the cross sections of the
process shown in \fig{ddhiggs} at LHC vary quite significantly
\cite{Bartels:2006ea,Gotsman:2008tr,Khoze:2007hx}. Two of the existing
approaches \cite{Gotsman:2008tr,Khoze:2007hx} were reviewed in the
talk by Kaidalov. Among other things he outlined the approximations
made for the triple pomeron vertex made in each of the approaches.
Both approaches reproduce the existing Tevatron double diffractive
data reasonably well, but differ significantly in their extrapolation
to LHC energies. Since it is not clear from first principles which
approximation of the triple pomeron vertex is better justified, it
seems that error analyses similar to those done for PDF's may be
needed to reconcile the differences between the two approaches in
question.

Physics beyond the Standard Model was discussed in the talks by V.V.
Khoze and Strassler dedicated to different supersymmetric models.
While the former talk presented a minimal approach to introducing
SUSY, the latter talk featured a broader range of possibilities. V.V.
Khoze talked about the ISS scenario \cite{Intriligator:2006dd} in
which the Universe lives in a metastable vacuum in which SUSY is
broken. At the same time there exists a hidden sector of the theory
with a true vacuum which is supersymmetric. The ISS model gives a
concrete example of SUSY breaking, allowing to calculate the mass
spectrum of the supersymmetric particles using the messenger fields.
Strassler in his talk argued that minimalistic approach to physics
beyond Standard Model is not necessarily what happens in nature and we
should prepare for big surprises at the LHC. He therefore talked about
hidden valleys and unparticles, both of which would lead to spectacular
hadronic shower events at the LHC, which unfortunately would be hard
to analyze and understand due to the large number of particles
produced. Indeed both minimal and non-minimal SUSY scenarios are quite
possible at LHC.

~\vspace*{0mm}

\section*{Acknowledgments}

I would like to acknowledge helpful discussions with Ulrich Heinz,
William Horowitz, Abhijit Majumder, and Heribert Weigert.

This work is sponsored in part by the U.S.  Department of Energy under
Grant No. DE-FG02-05ER41377.

\begin{footnotesize}

\begin{mcbibliography}{10}

\bibitem{JalilianMarian:2005jf}
J.~Jalilian-Marian and Y.~V. Kovchegov,
\newblock Prog. Part. Nucl. Phys.{} {\bf 56},~104~(2006).
\newblock \href{http://www.arXiv.org/abs/hep-ph/0505052}{{\tt
  hep-ph/0505052}}\relax
\relax
\bibitem{Dokshitzer:1977sg}
Y.~L. Dokshitzer,
\newblock Sov. Phys. JETP{} {\bf 46},~641~(1977)\relax
\relax
\bibitem{Gribov:1972ri}
V.~N. Gribov and L.~N. Lipatov,
\newblock Sov. J. Nucl. Phys.{} {\bf 15},~438~(1972)\relax
\relax
\bibitem{Altarelli:1977zs}
G.~Altarelli and G.~Parisi,
\newblock Nucl. Phys.{} {\bf B126},~298~(1977)\relax
\relax
\bibitem{Bal-Lip}
Y.~Y. Balitsky and L.~N. Lipatov,
\newblock Sov. J. Nucl. Phys.{} {\bf 28},~822~(1978)\relax
\relax
\bibitem{Kuraev:1977fs}
E.~A. Kuraev, L.~N. Lipatov, and V.~S. Fadin,
\newblock Sov. Phys. JETP{} {\bf 45},~199~(1977)\relax
\relax
\bibitem{Jalilian-Marian:1997gr}
J.~Jalilian-Marian, A.~Kovner, A.~Leonidov, and H.~Weigert,
\newblock Phys. Rev.{} {\bf D59},~014014~(1998).
\newblock \href{http://www.arXiv.org/abs/hep-ph/9706377}{{\tt
  hep-ph/9706377}}\relax
\relax
\bibitem{Iancu:2000hn}
E.~Iancu, A.~Leonidov, and L.~D. McLerran,
\newblock Nucl. Phys.{} {\bf A692},~583~(2001).
\newblock \href{http://www.arXiv.org/abs/hep-ph/0011241}{{\tt
  hep-ph/0011241}}\relax
\relax
\bibitem{Balitsky:1996ub}
I.~Balitsky,
\newblock Nucl. Phys.{} {\bf B463},~99~(1996).
\newblock \href{http://www.arXiv.org/abs/hep-ph/9509348}{{\tt
  hep-ph/9509348}}\relax
\relax
\bibitem{Kovchegov:1999yj}
Y.~V. Kovchegov,
\newblock Phys. Rev.{} {\bf D60},~034008~(1999).
\newblock \href{http://www.arXiv.org/abs/hep-ph/9901281}{{\tt
  hep-ph/9901281}}\relax
\relax
\bibitem{Lai:1999wy}
{ CTEQ} Collaboration, H.~L. Lai {\em et al.},
\newblock Eur. Phys. J.{} {\bf C12},~375~(2000).
\newblock \href{http://www.arXiv.org/abs/hep-ph/9903282}{{\tt
  hep-ph/9903282}}\relax
\relax
\bibitem{Martin:2007bv}
A.~D. Martin, W.~J. Stirling, R.~S. Thorne, and G.~Watt,
\newblock Phys. Lett.{} {\bf B652},~292~(2007).
\newblock \href{http://www.arXiv.org/abs/0706.0459}{{\tt 0706.0459}}\relax
\relax
\bibitem{Alekhin:2002fv}
S.~Alekhin,
\newblock Phys. Rev.{} {\bf D68},~014002~(2003).
\newblock \href{http://www.arXiv.org/abs/hep-ph/0211096}{{\tt
  hep-ph/0211096}}\relax
\relax
\bibitem{Ball:2008by}
{ NNPDF} Collaboration, R.~D. Ball {\em et al.}~(2008).
\newblock \href{http://www.arXiv.org/abs/0808.1231}{{\tt 0808.1231}}\relax
\relax
\bibitem{Salur:2008hs}
{ STAR} Collaboration, S.~Salur~(2008).
\newblock \href{http://www.arXiv.org/abs/0809.1609}{{\tt 0809.1609}}\relax
\relax
\bibitem{Adler:2003ii}
{ PHENIX} Collaboration, S.~S. Adler {\em et al.},
\newblock Phys. Rev. Lett.{} {\bf 91},~072303~(2003).
\newblock \href{http://www.arXiv.org/abs/nucl-ex/0306021}{{\tt
  nucl-ex/0306021}}\relax
\relax
\bibitem{Adams:2003im}
{ STAR} Collaboration, J.~Adams {\em et al.},
\newblock Phys. Rev. Lett.{} {\bf 91},~072304~(2003).
\newblock \href{http://www.arXiv.org/abs/nucl-ex/0306024}{{\tt
  nucl-ex/0306024}}\relax
\relax
\bibitem{Arsene:2003yk}
{ BRAHMS} Collaboration, I.~Arsene {\em et al.},
\newblock Phys. Rev. Lett.{} {\bf 91},~072305~(2003).
\newblock \href{http://www.arXiv.org/abs/nucl-ex/0307003}{{\tt
  nucl-ex/0307003}}\relax
\relax
\bibitem{Fadin:1998py}
V.~S. Fadin and L.~N. Lipatov,
\newblock Phys. Lett.{} {\bf B429},~127~(1998).
\newblock \href{http://www.arXiv.org/abs/hep-ph/9802290}{{\tt
  hep-ph/9802290}}\relax
\relax
\bibitem{Ciafaloni:1998gs}
M.~Ciafaloni and G.~Camici,
\newblock Phys. Lett.{} {\bf B430},~349~(1998).
\newblock \href{http://www.arXiv.org/abs/hep-ph/9803389}{{\tt
  hep-ph/9803389}}\relax
\relax
\bibitem{Andersen:2003wy}
J.~R. Andersen and A.~Sabio~Vera,
\newblock Nucl. Phys.{} {\bf B679},~345~(2004).
\newblock \href{http://www.arXiv.org/abs/hep-ph/0309331}{{\tt
  hep-ph/0309331}}\relax
\relax
\bibitem{Ciafaloni:2003rd}
M.~Ciafaloni, D.~Colferai, G.~P. Salam, and A.~M. Stasto,
\newblock Phys. Rev.{} {\bf D68},~114003~(2003).
\newblock \href{http://www.arXiv.org/abs/hep-ph/0307188}{{\tt
  hep-ph/0307188}}\relax
\relax
\bibitem{Altarelli:1999vw}
G.~Altarelli, R.~D. Ball, and S.~Forte,
\newblock Nucl. Phys.{} {\bf B575},~313~(2000).
\newblock \href{http://www.arXiv.org/abs/hep-ph/9911273}{{\tt
  hep-ph/9911273}}\relax
\relax
\bibitem{Kharzeev:2002pc}
D.~Kharzeev, E.~Levin, and L.~McLerran,
\newblock Phys. Lett.{} {\bf B561},~93~(2003).
\newblock \href{http://www.arXiv.org/abs/hep-ph/0210332}{{\tt
  hep-ph/0210332}}\relax
\relax
\bibitem{Kharzeev:2003wz}
D.~Kharzeev, Y.~V. Kovchegov, and K.~Tuchin,
\newblock Phys. Rev.{} {\bf D68},~094013~(2003).
\newblock \href{http://www.arXiv.org/abs/hep-ph/0307037}{{\tt
  hep-ph/0307037}}\relax
\relax
\bibitem{Albacete:2003iq}
J.~L. Albacete, N.~Armesto, A.~Kovner, C.~A. Salgado, and U.~A. Wiedemann,
\newblock Phys. Rev. Lett.{} {\bf 92},~082001~(2004).
\newblock \href{http://www.arXiv.org/abs/hep-ph/0307179}{{\tt
  hep-ph/0307179}}\relax
\relax
\bibitem{Arsene:2004ux}
{ BRAHMS} Collaboration, I.~Arsene {\em et al.},
\newblock Phys. Rev. Lett.{} {\bf 93},~242303~(2004).
\newblock \href{http://www.arXiv.org/abs/nucl-ex/0403005}{{\tt
  nucl-ex/0403005}}\relax
\relax
\bibitem{Adler:2004eh}
{ PHENIX} Collaboration, S.~S. Adler {\em et al.},
\newblock Phys. Rev. Lett.{} {\bf 94},~082302~(2005).
\newblock \href{http://www.arXiv.org/abs/nucl-ex/0411054}{{\tt
  nucl-ex/0411054}}\relax
\relax
\bibitem{Back:2004bq}
{ PHOBOS} Collaboration, B.~B. Back {\em et al.},
\newblock Phys. Rev.{} {\bf C70},~061901~(2004).
\newblock \href{http://www.arXiv.org/abs/nucl-ex/0406017}{{\tt
  nucl-ex/0406017}}\relax
\relax
\bibitem{Adams:2006uz}
{ STAR} Collaboration, J.~Adams {\em et al.},
\newblock Phys. Rev. Lett.{} {\bf 97},~152302~(2006).
\newblock \href{http://www.arXiv.org/abs/nucl-ex/0602011}{{\tt
  nucl-ex/0602011}}\relax
\relax
\bibitem{Eskola:2008ca}
K.~J. Eskola, H.~Paukkunen, and C.~A. Salgado,
\newblock JHEP{} {\bf 07},~102~(2008).
\newblock \href{http://www.arXiv.org/abs/0802.0139}{{\tt 0802.0139}}\relax
\relax
\bibitem{Kovchegov:2001sc}
Y.~V. Kovchegov and K.~Tuchin,
\newblock Phys. Rev.{} {\bf D65},~074026~(2002).
\newblock \href{http://www.arXiv.org/abs/hep-ph/0111362}{{\tt
  hep-ph/0111362}}\relax
\relax
\bibitem{Kharzeev:2001yq}
D.~Kharzeev, E.~Levin, and M.~Nardi,
\newblock Phys. Rev.{} {\bf C71},~054903~(2005).
\newblock \href{http://www.arXiv.org/abs/hep-ph/0111315}{{\tt
  hep-ph/0111315}}\relax
\relax
\bibitem{Albacete:2007sm}
J.~L. Albacete,
\newblock Phys. Rev. Lett.{} {\bf 99},~262301~(2007).
\newblock \href{http://www.arXiv.org/abs/0707.2545}{{\tt 0707.2545}}\relax
\relax
\bibitem{Kovchegov:2006vj}
Y.~V. Kovchegov and H.~Weigert,
\newblock Nucl. Phys.{} {\bf A784},~188~(2007).
\newblock \href{http://www.arXiv.org/abs/hep-ph/0609090}{{\tt
  hep-ph/0609090}}\relax
\relax
\bibitem{Balitsky:2006wa}
I.~Balitsky,
\newblock Phys. Rev.{} {\bf D75},~014001~(2007).
\newblock \href{http://www.arXiv.org/abs/hep-ph/0609105}{{\tt
  hep-ph/0609105}}\relax
\relax
\bibitem{Albacete:2007yr}
J.~L. Albacete and Y.~V. Kovchegov,
\newblock Phys. Rev.{} {\bf D75},~125021~(2007).
\newblock \href{http://www.arXiv.org/abs/0704.0612}{{\tt 0704.0612}}\relax
\relax
\bibitem{Daugherity:2008su}
{ STAR} Collaboration, M.~Daugherity,
\newblock J. Phys.{} {\bf G35},~104090~(2008).
\newblock \href{http://www.arXiv.org/abs/0806.2121}{{\tt 0806.2121}}\relax
\relax
\bibitem{CasalderreySolana:2004qm}
J.~Casalderrey-Solana, E.~V. Shuryak, and D.~Teaney,
\newblock J. Phys. Conf. Ser.{} {\bf 27},~22~(2005).
\newblock \href{http://www.arXiv.org/abs/hep-ph/0411315}{{\tt
  hep-ph/0411315}}\relax
\relax
\bibitem{Adare:2007vu}
{ PHENIX} Collaboration, A.~Adare {\em et al.},
\newblock Phys. Rev.{} {\bf C77},~011901~(2008).
\newblock \href{http://www.arXiv.org/abs/0705.3238}{{\tt 0705.3238}}\relax
\relax
\bibitem{Kovner:2002yt}
A.~Kovner and U.~A. Wiedemann,
\newblock Phys. Lett.{} {\bf B551},~311~(2003).
\newblock \href{http://www.arXiv.org/abs/hep-ph/0207335}{{\tt
  hep-ph/0207335}}\relax
\relax
\bibitem{Lappi:2006fp}
T.~Lappi and L.~McLerran,
\newblock Nucl. Phys.{} {\bf A772},~200~(2006).
\newblock \href{http://www.arXiv.org/abs/hep-ph/0602189}{{\tt
  hep-ph/0602189}}\relax
\relax
\bibitem{Mrowczynski:1993qm}
S.~Mrowczynski,
\newblock Phys. Lett.{} {\bf B314},~118~(1993)\relax
\relax
\bibitem{Arnold:2003rq}
P.~Arnold, J.~Lenaghan, and G.~D. Moore,
\newblock JHEP{} {\bf 08},~002~(2003).
\newblock \href{http://www.arXiv.org/abs/hep-ph/0307325}{{\tt
  hep-ph/0307325}}\relax
\relax
\bibitem{Kovner:1995ja}
A.~Kovner, L.~D. McLerran, and H.~Weigert,
\newblock Phys. Rev.{} {\bf D52},~6231~(1995).
\newblock \href{http://www.arXiv.org/abs/hep-ph/9502289}{{\tt
  hep-ph/9502289}}\relax
\relax
\bibitem{Kovchegov:1997ke}
Y.~V. Kovchegov and D.~H. Rischke,
\newblock Phys. Rev.{} {\bf C56},~1084~(1997).
\newblock \href{http://www.arXiv.org/abs/hep-ph/9704201}{{\tt
  hep-ph/9704201}}\relax
\relax
\bibitem{Krasnitz:2003jw}
A.~Krasnitz, Y.~Nara, and R.~Venugopalan,
\newblock Nucl. Phys.{} {\bf A727},~427~(2003).
\newblock \href{http://www.arXiv.org/abs/hep-ph/0305112}{{\tt
  hep-ph/0305112}}\relax
\relax
\bibitem{Kovchegov:2007vf}
Y.~V. Kovchegov and H.~Weigert,
\newblock Nucl. Phys.{} {\bf A807},~158~(2008).
\newblock \href{http://www.arXiv.org/abs/0712.3732}{{\tt 0712.3732}}\relax
\relax
\bibitem{Maldacena:1997re}
J.~M. Maldacena,
\newblock Adv. Theor. Math. Phys.{} {\bf 2},~231~(1998).
\newblock \href{http://www.arXiv.org/abs/hep-th/9711200}{{\tt
  hep-th/9711200}}\relax
\relax
\bibitem{Witten:1998qj}
E.~Witten,
\newblock Adv. Theor. Math. Phys.{} {\bf 2},~253~(1998).
\newblock \href{http://www.arXiv.org/abs/hep-th/9802150}{{\tt
  hep-th/9802150}}\relax
\relax
\bibitem{Janik:2005zt}
R.~A. Janik and R.~B. Peschanski,
\newblock Phys. Rev.{} {\bf D73},~045013~(2006).
\newblock \href{http://www.arXiv.org/abs/hep-th/0512162}{{\tt
  hep-th/0512162}}\relax
\relax
\bibitem{Bjorken:1982qr}
J.~D. Bjorken,
\newblock Phys. Rev.{} {\bf D27},~140~(1983)\relax
\relax
\bibitem{Landau:1953gs}
L.~D. Landau,
\newblock Izv. Akad. Nauk SSSR Ser. Fiz.{} {\bf 17},~51~(1953)\relax
\relax
\bibitem{Bearden:2003hx}
{ BRAHMS} Collaboration, I.~G. Bearden {\em et al.},
\newblock Phys. Rev. Lett.{} {\bf 93},~102301~(2004).
\newblock \href{http://www.arXiv.org/abs/nucl-ex/0312023}{{\tt
  nucl-ex/0312023}}\relax
\relax
\bibitem{Huovinen:2001cy}
P.~Huovinen, P.~F. Kolb, U.~W. Heinz, P.~V. Ruuskanen, and S.~A. Voloshin,
\newblock Phys. Lett.{} {\bf B503},~58~(2001).
\newblock \href{http://www.arXiv.org/abs/hep-ph/0101136}{{\tt
  hep-ph/0101136}}\relax
\relax
\bibitem{Teaney:2001av}
D.~Teaney, J.~Lauret, and E.~V. Shuryak~(2001).
\newblock \href{http://www.arXiv.org/abs/nucl-th/0110037}{{\tt
  nucl-th/0110037}}\relax
\relax
\bibitem{Policastro:2001yc}
G.~Policastro, D.~T. Son, and A.~O. Starinets,
\newblock Phys. Rev. Lett.{} {\bf 87},~081601~(2001).
\newblock \href{http://www.arXiv.org/abs/hep-th/0104066}{{\tt
  hep-th/0104066}}\relax
\relax
\bibitem{Kovtun:2004de}
P.~Kovtun, D.~T. Son, and A.~O. Starinets,
\newblock Phys. Rev. Lett.{} {\bf 94},~111601~(2005).
\newblock \href{http://www.arXiv.org/abs/hep-th/0405231}{{\tt
  hep-th/0405231}}\relax
\relax
\bibitem{Lisa:2008gf}
M.~A. Lisa and S.~Pratt~(2008).
\newblock \href{http://www.arXiv.org/abs/0811.1352}{{\tt 0811.1352}}\relax
\relax
\bibitem{Kovchegov:2007pq}
Y.~V. Kovchegov and A.~Taliotis,
\newblock Phys. Rev.{} {\bf C76},~014905~(2007).
\newblock \href{http://www.arXiv.org/abs/0705.1234}{{\tt 0705.1234}}\relax
\relax
\bibitem{Aharony:1999ti}
O.~Aharony, S.~S. Gubser, J.~M. Maldacena, H.~Ooguri, and Y.~Oz,
\newblock Phys. Rept.{} {\bf 323},~183~(2000).
\newblock \href{http://www.arXiv.org/abs/hep-th/9905111}{{\tt
  hep-th/9905111}}\relax
\relax
\bibitem{Albacete:2008ze}
J.~L. Albacete, Y.~V. Kovchegov, and A.~Taliotis,
\newblock JHEP{} {\bf 07},~074~(2008).
\newblock \href{http://www.arXiv.org/abs/0806.1484}{{\tt 0806.1484}}\relax
\relax
\bibitem{Dominguez:2008vd}
F.~Dominguez, C.~Marquet, A.~H. Mueller, B.~Wu, and B.-W. Xiao,
\newblock Nucl. Phys.{} {\bf A811},~197~(2008).
\newblock \href{http://www.arXiv.org/abs/0803.3234}{{\tt 0803.3234}}\relax
\relax
\bibitem{McLerran:1993ni}
L.~D. McLerran and R.~Venugopalan,
\newblock Phys. Rev.{} {\bf D49},~2233~(1994).
\newblock \href{http://www.arXiv.org/abs/hep-ph/9309289}{{\tt
  hep-ph/9309289}}\relax
\relax
\bibitem{Bern:2005iz}
Z.~Bern, L.~J. Dixon, and V.~A. Smirnov,
\newblock Phys. Rev.{} {\bf D72},~085001~(2005).
\newblock \href{http://www.arXiv.org/abs/hep-th/0505205}{{\tt
  hep-th/0505205}}\relax
\relax
\bibitem{Bartels:2006ea}
J.~Bartels, S.~Bondarenko, K.~Kutak, and L.~Motyka,
\newblock Phys. Rev.{} {\bf D73},~093004~(2006).
\newblock \href{http://www.arXiv.org/abs/hep-ph/0601128}{{\tt
  hep-ph/0601128}}\relax
\relax
\bibitem{Gotsman:2008tr}
E.~Gotsman, E.~Levin, U.~Maor, and J.~S. Miller~(2008).
\newblock \href{http://www.arXiv.org/abs/0805.2799}{{\tt 0805.2799}}\relax
\relax
\bibitem{Khoze:2007hx}
V.~A. Khoze, A.~D. Martin, and M.~G. Ryskin,
\newblock Phys. Lett.{} {\bf B650},~41~(2007).
\newblock \href{http://www.arXiv.org/abs/hep-ph/0702213}{{\tt
  hep-ph/0702213}}\relax
\relax
\bibitem{Intriligator:2006dd}
K.~A. Intriligator, N.~Seiberg, and D.~Shih,
\newblock JHEP{} {\bf 04},~021~(2006).
\newblock \href{http://www.arXiv.org/abs/hep-th/0602239}{{\tt
  hep-th/0602239}}\relax
\relax
\end{mcbibliography}
\providecommand{\etal}{et al.\xspace}
\providecommand{\href}[2]{#2}
\providecommand{\coll}{Coll.}
\catcode`\@=11
\def\@bibitem#1{%
\ifmc@bstsupport
  \mc@iftail{#1}%
    {;\newline\ignorespaces}%
    {\ifmc@first\else.\fi\orig@bibitem{#1}}
  \mc@firstfalse
\else
  \mc@iftail{#1}%
    {\ignorespaces}%
    {\orig@bibitem{#1}}%
\fi}%
\catcode`\@=12

\end{footnotesize}

\end{document}